# The use of biophysical approaches to understand ciliary beating

Pietro Cicuta

Cavendish Laboratory

University of Cambridge, Cambridge CB3 0HE, U.K.

Pc245@cam.ac.uk

**Abstract**

Motile cilia are a striking example of functional cellular organelle, conserved across all the eukaryotic species. Motile cilia allow swimming of cells and small organisms and transport of liquids across epithelial tissues. Whilst the molecular structure is now very well understood, the dynamics of cilia is not well established either at the single cilium level nor at the level of collective beating. Indeed, a full understanding of this requires connecting together behaviour across various lengthscales, from the molecular to the organelle, then at cellular level and up to the tissue scale. Aside from the fundamental interest in this system, understanding beating is important to elucidate aspects of embryonic development and a variety of health conditions from fertility to genetic and infectious diseases of the airways.

**Introduction**

Thanks to the work of the last 70 years we have a complete picture of the structure of many organelles in a cell, in terms of the constituent proteins, and also in many cases a good sense of the regulatory and metabolic molecules involved in their function.  And yet, particularly in the cases of dynamical processes, we struggle (apart from few examples) to understand how all these components come together into robust phenotypes [1]. In the tradition of physics, there are many examples non-living systems with such `emergent properties', for example phase transitions of matter (solid/fluid/gas equilibrium states) or synchronisation of pendula. What makes these very different systems similar to each other and can shed light perhaps on a variety of biological phenotypes is the fact that we can describe them all with predictive models, in situations where the `rules' acting on one level of time and length scales do not `encode' at all any of the behaviour at the scale of the phenotype. The phenotype is the `emergent phenomenon'. Of course this happens all the time at the level of basic biochemistry: the chemical interactions of macromolecules provide the network of reactions that supports pathways and metabolism. However in cells and tissues much of this biochemistry is also regulated and feeds back into physical interactions. A classical example of biochemistry connected to physical phenomena  is the voltage gating of ion channels, and another is the role of mechanical forces on cell polarisation. Many of the most poorly understood processes in biology seem to be ones that connect strongly quite distinct pieces of science, for example physical forces affecting gene regulation and cell differentiation [2] or thermodynamical conditions affecting protein interactions [3].

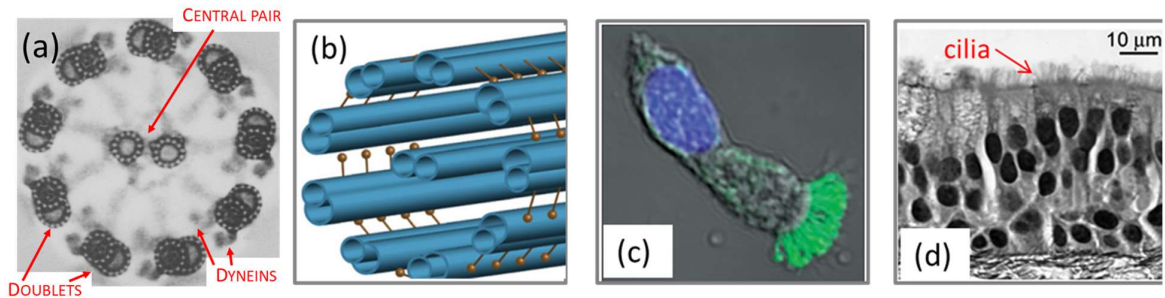

Figure 1. **Overview of the multi-scale structures at play in motile cilia.** The cilia structure and the anatomy of ciliated epithelia are well known, but how these modules give rise to well defined beating dynamics is still poorly understood. (a) Electron micrograph of a motile cilium cross section, showing the 9+2 microtubule ultrastructure, and dynein arms [48]. The cilia are approximately 200nm in diameter. (b) Schematic of the ultrastructure, highlighting the dynein motors that run all along the axoneme [49]. At this scale, models have been proposed [14] to describe the cooperative binding/unbinding of motors, to give rise to a waveform on the cilia. (c) Fluorescence optical microscopy of a single cell from respiratory tract, with cilia labelled in green and nucleus in blue [18]. These cilia are approximately 10µm in length. (d) Stained cross-section of an epithelial cell layer from culture, showing the cilia "carpet" at the cell/fluid interface [37].

Fluid dynamics is a well developed body of knowledge that is very relevant in a variety of questions in living matter [4] [5] [6]. In most processes at the cell or tissue scale the flows are laminar, and one has to understand velocity fields, to calculate the forces that a fluid can carry from one solid boundary to another. Laminar flow fields are simpler than the general case of flow, but unfortunately biological systems typically add a great deal of complexity through the fluid composition. Most biological fluids are quite concentrated macromolecular solutions, with flow properties that depend on the rate of deformation in a complex fashion.

Motile cilia are cellular organelles that beat periodically in the liquid surrounding a cell, or adjacent to a ciliated epithelium. Their whip-like movement has evolved to quite efficiently displace fluid, allowing either cell propulsion (e.g. sperm, *Paramecium* and *Chlamydomonas*) or transport of liquid across an epithelium. Cilia beating is a biological phenomenon conserved from unicellular to multicellular eukaryotes, including cells of plants and vertebrates. In motile cilia, dynein molecular motors hydrolyse ATP to exert force on microtubules, causing the cilia to bend periodically. The ultrastructure, the "axoneme", of motile cilia is almost always made of nine microtubule doublets and two central singlet tubes, known as `9+2', as shown in Figure 1(a) (we only know of the left-right organiser cilium in vertebrates as an example of a motile cilium lacking the central pair). At this scale, there is already an example of emergent phenomenon: the hundreds of dynein motors along the cilium (see Figure 1(b)) act cooperatively in order to bend the cilium. One motor would not in itself have enough force to bend the structure. Various theories have been put forward for how the system of motors + cilium functions, but there is still debate [7] in how to describe this. Video microscopy of individual cilia shows that the cilium bends to very much the same shapes periodically in time, usually with frequencies between 5 and 50 Hz.

This beating is in fact a traveling wave, down the cilium. If one "zooms out", and looks at multiciliated cells, then another phenomenon is observed: the cilia beat in synchrony to each other, in phase or sometimes with a small but constant phase difference. In a ciliated epithelium, this property is conserved across many cells and is crucial to allow fluid to be pushed very efficiently in a given direction (which is the direction of planar cell polarity) over scales of centimeters or more. This coordination of cilia is again an emergent property. In this review we will provide some background at the single cilium level, but mostly focus on the work of ourselves and others in the last 10 years on the question of cilia synchronisation. For the correct function of various human organs, cilia have to beat, but also they have to beat in such a way that they synchronise to transport fluids.

**Motile cilia as multi-scale systems**

- *Structure at the molecular scale*

Cilia structure is cell biology textbook material [1] [8]: all motile cilia in nature have 9 microtubule doublets and 2 single central microtubules, see Figure 1(a). The proteins linking these tubules to each other are well known, as is the centriolar structure that lies at the base of the cilium. Activity in the motile cilia arises from the molecular motor dynein (structural differences across species), which binds across the microtubule doublets. Dynein hydrolyses ATP to produce a stroke, which acts to slide one doublet versus the other. In connection with the constraints posed by various bridging structures, this results in a temporally and spatially localised bending of the microtubules, which is in turn resisted by the elastic rigidity of the tubules. Calcium is well known to regulate cilia activity.

- *Beating at the cilium scale – – calcium, ATP consumption, efficiency of beat. How does the structure "turn" into a robust and well defined dynamics for the cilia waveform.*

It is still an open question how to make the link from structure and basic activity introduced above, to a well defined, robust and periodic beating of the cilium, along its full length of several microns [9]. Each cilium undergoes periodically a forward "power stroke", followed by a backward "recovery stroke" in which it is more bent and closer to the surface, thus resulting in a net momentum transfer to the fluid over a cycle. It is clear from observations that the cilium bends through a wave of curvature, which propagates from one end, usually the base, to the tip [10]. This implies that the localised events of each dynein motor are coupled to each other. Recent advances in electron microscopy have provided snapshots of the motor configurations, in connection to the filament curvature [11] [12]. To understand how this all fits together as a steady state of dynamics, one needs to know how dyneins on opposing sides of the axoneme can take turns binding and exerting force, and then how this mechanism can be extended to describe the spatial propagation along the axoneme. The first proposals in this direction are reviewed in [9] [13], as are a number of investigations on various species. Conceptually there are really only two possible sources for the coordination of motors: the dyneins could feel local configuration, or they could feel the local force. Configuration and force are coupled together, since bending requires a force and viceversa, and yet these scenarios are fundamentally different and it's important to reach a conclusion as to which way the causality is acting. We need this in order to understand how the beat is subtly regulated across cell types and species, and how a cilium beat will respond to external forces, react to changes in the medium viscoelasticity, and be robust to sources of noise. A recent model by Julicher and co-workers [14] is an example of how to capture the complex waveform into a simple model that retains key physical elements, and such models can be built upon a force – or configuration-feedback. Despite very nice work that goes a long way to explaining the propagating waves on cilia [7] we are not aware of a conclusive experiment on this question of how to represent the basic element of active force. Such an experiment cannot just be based on detailed imaging of configurations, it would need a study of how the waveform reacts to external forcing.

- *The scale across cilia – the question of synchronisation, and travelling waves. Essential for mucociliary clearance, flow in the brain, fallopian tube transport.*

Our own interest in motile cilia (Figure 2) is principally to understand the physical principles of how two or more cilia manage to coordinate their beating, typically phase-locking their dynamics. This question has important consequences in human health, especially in respiratory physiology and disease, where motile cilia play a vital role on the surface of the airways, maintaining an upward flow of mucus, away from the lungs [15] [16]. They also determine the asymmetry of various organisms during development [17]. A range of pathologies are associated to defects of motile cilia [18].

In arrays of cilia the beating is synchronised, generating complex wave-like patterns called metachronal waves [19] [20]. The simplest biological model for this question of inter-cilia interactions is the alga *Chlamydomonas reinhardii,* which has two cilia (note that eukaryotic cilia are called `flagella' in some systems, which is confusing because the term also refers to the completely unrelated filaments of bacteria; in eukaryotes cilia and flagella are synonymous). Recent very clear experiments have explored the phenomenology of synchronised flagella in the simple algae systems [21] [22] [23], showing how cilia are mechanically coupled by the flow of fluid, typically in the low Reynolds number (Re) regime [4]. These studies opened up an area of quantitative modelling with statistical physics and fluid dynamics.

Our own work in this direction started with artificial systems, understanding the role of fluid flow by creating phase-free oscillators with colloidal particles and optical trapping [24]. It is possible to represent the fluid flow inducted by a cilium, by moving a point force (physically, a colloidal sized sphere) around a closed trajectory; the forces and velocities as a function of the beating cycle can be mapped to those of a biological system, by moving the particle around the `center of drag' [25]. These quite abstract studies highlighted two key facts: (a) hydrodynamic flows of magnitude that cilia can exert are sufficient to cause phase locking [26] [27]; (b) the importance of the ciliary waveform (which is mimicked, in the artificial system, by the way a particle is forced in the periodic motion) - with small modulations of the way each cilium beats, one can get completely different collective dynamics [28] [29] [25]. Figure 2 illustrates these studies by showing the pipelines of analysis: quantitative data can be obtained at the scale of the cilium (a) or across cilia (d,e), and it is possible to create models (illustrated in panels (b) and (c)) to try and link the behaviour at the cilium scale to the behaviour across cilia.

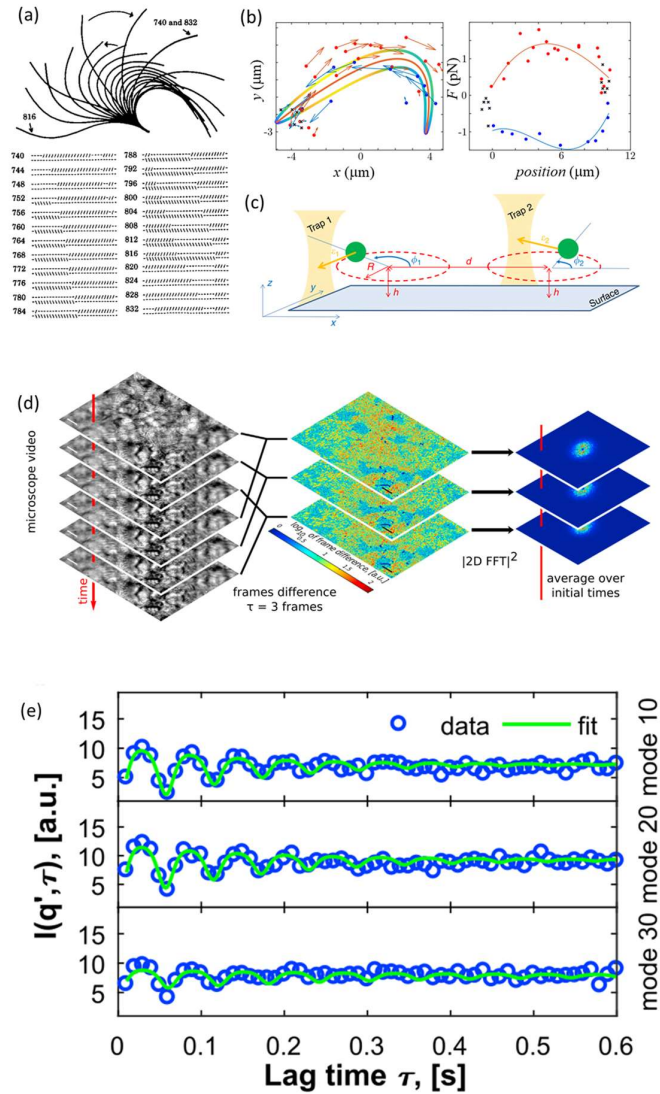

Figure 2. **There are two scales at which emergent collective dynamics manifests in motile cilia, and the dynamics can be measured quantitatively at both scales.** At the level of individual cilia, the dynein molecular motors synchronise their attachment/detachment, in a way that leads to cyclic power/recovery strokes of the cilium, which we call the cilium waveform. At the scale across cilia, there is collective dynamics leading physiologically to travelling phase waves called metachronal waves. (a) A computer simulation of the first "geometric clutch" model addressing emergence of the waveform [47]. The waveform from biological systems can be tracked from timelapse videos, and (b) shows as an example the position of the center of drag and the corresponding force over the beat cycle [33]. The data from (b) can be used to understand how to cilia will synchronise due to the fluid flow they exert on each other, using simple models in experiment (c) or simulation [25]. From timelapse of epithelia (d), using multiDDM without any image analysis steps one can measure oscillatory signals (e) which contain the distribution of beat frequencies but also the length across which cilia dynamics is synchronised [42].

**Experimental approaches**

- *In vitro and in silico approaches, and model organisms.*

It is possible to isolate the axonemes from the cell body, for example from Chlamydomonas, and perform in-vitro assays of the cilia motility.   These assays allow great optical access, and can provide insight into ATP consumption, the role of calcium, and the action of drugs or toxins. However one cannot expect these waveforms to be the same as in physiological conditions. Since the base is not anchored, the boundary condition on the cilia base is very different (this "free" base can move and twist), and also the cilia are typically on a surface which changes their fluid resistive forces and drag. So assays in these conditions are probably less useful to assess the detailed shapes of the waveform.

Simulations are not yet powerful enough to attempt an atomistic representation of ciliary dynamics (at the moment it is possible to simulate one stroke of a molecular motor, so perhaps in a few years' time we will have a simulation of a detailed axoneme).  At the moment, the power of simulations lies in running simpler models, both at the scale of one cilium [14] [30] [31], and at the scale of cilia interacting with each other [32] [25] [33].

Algae cells have proven a very fruitful experimental platform in the last decade [5]. *Chlamydomonas* cells, and also somatic cells from the colony organism *Volvox*, can be held in micropipettes and observed in optical microscopy, giving detailed cilia waveforms and enabling recording of phase locking.  Similarly, sperm cells from a variety of organisms have been studied for a long time: the travelling wave on their axoneme is very distinct, mostly due to the difference in base coupling, the long axoneme, and the external membrane covering this cilium [34]. The embryos of *xenopus* have been a model system for investigating the developmental biology of ciliated epithelia [35].  The single cell organism *paramecium* uses collective beating of hundreds of cilia on its surface to propel itself, in a striking and classical example of cilia metachronal waves [36]. All of these model organisms are in different ways being used to study human-related genetic ciliopathies.

- *Cilia experiments with ALI and respiratory epithelia*

Understanding the physiology and clinically-related problems of the respiratory tract is probably the main driver of research in motile cilia. The muco-ciliated epithelium in the airways is the first line of defence against a wide range of potentially harmful agents. The layer of mucus (ions, water, mucin proteins), kept in motion by beating cilia and occasionally by cough, acts as a physical barrier. Flow of this fluid is maintained over all the airway surfaces, toward the larynx. As reviewed in [37] dysfunction in this clearing system has severe consequences such as repeated lung infections and respiratory insufficiency. Mucociliary clearance is compromised in a variety of conditions, both genetic like cystic fibrosis (CF), primary ciliary dyskinesia (PCD), or acquired like chronic obstructive pulmonary disease (COPD). Altered mucus clearance is also linked to other chronic airway diseases such as asthma that afflict millions worldwide. The human bronchial system is complex: epithelium is mucociliated and pseudo-stratified in vivo, and the mechanisms that regulate its functions are not fully understood. Among epithelial cell layers, goblet cells secrete mucus, which protects the bronchial epithelium and is transported by the ciliary beats at the apical pole of specialised epithelial cells. The coordinated system of epithelial cells -including functions as ion transport through their membranes, mucus secretions and cilia action- and cough, is collectively termed mucus clearance, and results in a continuous flow.

The respiratory tract is lined by a thin layer of airway surface liquid (ASL), which is approximately 7-70 µm in height [38]. This layer is itself defined as consisting of two phases: an outer mucus-rich layer, and an inner fluid close to the cilia, known as "periciliary liquid layer" (PCL). The PCL height is in practice defined as height of the extended cilium, and recent studies have demonstrated complex interactions among cilia, the mobile mucins in the mucus layer, and the underlying PCL. The flow in the boundary PCL layer is clearly very complex, and has been addressed from a colloidal model perspective by [39].

I am not aware of endoscopy assays able to assess cilia motion in-situ in humans. Experiments on mucociliary clearance can be carried out on biopsies, by recording the displacements of tracer particles added to the epithelium. The motion of tracer particles is straightforward to track or quantify via Particle Image Velocimetry (PIV). This is a valuable approach in research, using animal models [40]. In clinical studies, it is common to diagnose cilia function by optical microscopy, having scraped cells from the nose or upper airways. Traditionally a clinician with experience on what constitutes a range of healthy or problematic waveforms would classify the dynamics, but recently there have been reports showing that this approach could be automated [10] [41].

Through a process of de-differentiation, expansion and differentiation back to ciliated epithelia, the primary cells from a nose scrape can be regrown to a confluent ciliated epithelium. To ciliate, airway cells need to be exposed to air, so a technique known as Air-Liquid interface culture (ALI) has been developed. There are commercially available petridish inserts to maintain the cells in culture medium on one side, and exposed to air on the other. These ALI cultures are time consuming (several weeks of culture) but allow a variety of experiments, and they are accessible to optical microscopy. They are particularly useful in the study of genetic conditions, but also in drug testing and to understand questions of developmental biology. Cilia from an ALI culture can be investigated to observe the waveform, e.g. [10].

Our team has recently demonstrated a new video analysis technique we called multi-DDM to extract information from high frame-rate videos of cells in ALI culture [42] [43]. The basic property of cilia beating is their frequency (CBF) but a wealth of other information is present in videos of cells at ALI, and connected to the fundamental concepts and open questions introduced earlier. The multi-DDM approach allows one to extract: (1) the CBF across the tissue; (2) the spatial decay of correlation in frequency, i.e. how coherent are neighbouring cilia, and how is this coherence maintained over distance; (3) the temporal coherence of each cilia; (4) the correlation of phase between neighbouring cilia.

A new and very exciting area of investigation is emerging around experiments where a perturbation is applied on the motile cilia. As described above, this is likely to be essential in understanding what is the underlying correct physical mechanism to describe the emergent properties. By applying external forces (typically oscillatory flows, of controlled frequency and velocity) we can hope to understand better both the emergence of waveforms within the cilium, and the emergence of collective dynamics. The first experiments in this direction were done on *Chlamydomonas* [44] [45] and pointed at the importance of cytoskeletal coupling to support synchronisation of cilia beat. The implication suggested by the authors is that the cytoskeleton provides a physical elastic link to transmit forces across cilia. We recently explored this question numerically, with data across species [33], and we believe that the experiments in [44] should be interpreted differently for two reasons: (1) the detailed shape of the waveform needs to be considered, in interpreting the results; (2) the waveform could change on modifying anchoring to cytoskeleton. But the fact remains that external flow experiments are necessary to probe these systems at the level required to distinguish mechanisms. Indeed we built on the experiment of [44] to apply oscillatory flows to multiciliated cells of mammalian brain [46], and the evidence from this study is that hydrodynamic coupling is sufficient to explain collective dynamics in that ciliated epithelium.

**Focus on what physical aspects allow cilia to beat cooperatively**

The physical methods, both experimental and modelling, applied to cilia motility and particularly to the question of collective dynamics have shown the importance first of all of geometric properties in the tissue, like distance and alignment of cilia. In cases where hydrodynamic interaction dominates, these parameters set the strength of interaction. In these systems also the kinematic properties of the fluid (viscosity, and, if the fluid is viscoelastic, it's full rheological behaviour) are crucial. We saw clear evidence of this in experiments where the cilia frequency but also the range of coordination changed dramatically with mucus washing and drug treatment, in CF cells [43].

A less obvious effect is due to the cilia waveforms. Even if the cilia are close together, and aligned, it is possible for them to have very little influence on each other. This will be the case, in practice, for rigid cilia undergoing similar power/recovery strokes, a situation seen in some cases of PCD [10]. In our study of waveforms across species [33] we explain how there is a trade off between pushing fluid at the level of individual cilia beats, and having a waveform that will give strong synchronisation – motile cilia from organisms and systems where one of the two functions dominates clearly classify into one or other of the types of waveform.

**Perspectives**

(i) *Importance of the field.* Motile cilia are an organelle expressed in many eukaryotes, with a conserved structure. Motile cilia underpin a variety of biological processes. In human health they are implicated in fertility, embryo development and physiology of the airways.

(ii) *Current thinking.* We do not yet have an accepted model neither to explain cilia waveforms nor collective dynamics, in a way that can be simulated numerically with predictive value. We are also currently unable to generate in vitro cell cultures with the same degree of cilia orientation and density as a human tissue, which limits personalised medicine and tissue transplants.

(iii) *Future directions.* We urgently need to resolve the two fundamental questions of how motile cilia beat: the level of waveform, i.e. what combination of force and configuration are motors reacting to, and how they influence each other, whether hydrodynamics or cytoskeletal coupling both need to be accounted for. The first question is likely to be universal, whereas the second might be species dependent. Both questions are likely to be resolved through careful experiment and analysis following perturbations with applied flow. This understanding will give robust underpinning for large scale simulations that can include fluid dynamics, geometry, etc. in ways that will be realistic even in complex systems like the airways. In terms of experimental conditions, we expect organoid cultures to reproduce ever more physiological conditions and shed light on developmental biology aspects.

**Conflicts of interest, acknowledgements, funding information**

The author has no conflicts of interest. I wish to acknowledge all my collaborators over the last 10 years who helped make our work possible, and particularly N.Bruot, J.Kotar, E.Hamilton, L.Feriani and N.Pellicciotta who worked in my lab developing the physical concepts reviewed here. This work was the focus of ERC CoG grant HydroSync.